# Controls and Machine Protection Systems

*E. Carrone*
SLAC National Accelerator Laboratory, Menlo Park, CA, USA

**Abstract**
Machine protection, as part of accelerator control systems, can be managed with a 'functional safety' approach, which takes into account product life cycle, processes, quality, industrial standards and cybersafety. This paper will discuss strategies to manage such complexity and the related risks, with particular attention to fail-safe design and safety integrity levels, software and hardware standards, testing, and verification philosophy. It will also discuss an implementation of a machine protection system at the SLAC National Accelerator Laboratory's Linac Coherent Light Source (LCLS).

**Keywords**
MPS; Functional Safety; PLC; SIL; Control Systems; Cyber Security.

## 1       A software problem

On 4 June 1996, the maiden flight of the Ariane 5 launcher ended in a failure. Only 39 s after initiation of the flight sequence, at an altitude of about 3700 m, the launcher veered off its flight path, broke up, and exploded.

During those first 39 s, the software generated a number too large for the system to handle: the computer shut down and passed control to its redundant twin, which, being identical to the first, came to the same conclusion and shut down a few milliseconds later. The rocket, now without guidance, changed direction to compensate for an imagined error and collapsed in its own turbulence.

In general terms, the flight control system of the Ariane 5 is of a standard design. The attitude of the launcher and its movements in space are measured by an inertial reference system. It has its own internal computer, in which angles and velocities are calculated on the basis of information from an inertial platform, with laser gyroscopes and accelerometers. The data from the inertial reference system are transmitted through the databus to the onboard computer, which executes the flight program and controls the nozzles of the solid boosters and the Vulcain cryogenic engine, via servo valves and hydraulic actuators.

To improve the reliability of such a system, there is considerable redundancy at the equipment level: two inertial reference systems operate in parallel, with identical hardware and software. One inertial reference system is active and one is in 'hot' standby; if the onboard computer detects that the active inertial reference system has failed, it immediately switches to the other one, provided that this unit is functioning properly. Likewise, there are two onboard computers, and a number of other units in the flight control system are also duplicated.

The launcher started to disintegrate at about 39 s into operation because of high aerodynamic loads due to an angle of attack of more than 20° that led to separation of the boosters from the main stage, in turn triggering the self-destruct system of the launcher. This angle of attack was caused by full nozzle deflections of the solid boosters and the main engine.

These nozzle deflections were commanded by the onboard computer software on the basis of data transmitted by the active inertial reference system. Part of these data at that time did not contain proper flight data, but showed a diagnostic bit pattern of the computer of the inertial reference system 2, which was interpreted as flight data. The reason that the active inertial reference system 2 did not send correct attitude data was that the unit had declared a failure due to a software exception.

The onboard computer could not switch to the back-up inertial reference system 1 because that unit had already ceased to function during the previous data cycle (72 ms) for the same reason as inertial reference system 2.

The internal inertial reference system software exception was caused during execution of a data conversion from 64-bit floating point to 16-bit signed integer value. The floating point number that was converted had a value greater than could be represented by a 16-bit signed integer. This resulted in an operand error.

Among the causes:

— software reused from the Ariane 4 series (a rocket with different requirements);

— an error while converting a 64-bit floating point number to a 16-bit integer caused an overflow, a custom floating point format for which the processor could have generated an exception error;

— some operations (in Ada code) on the computers are protected from bad conversions, but one was disabled;

— the primary inertial sub-computer and its back-up both shut down because of this, and the primary sub-computer started a memory dump;

— the main computer looked at the data dump and interpreted it as flight data. The nozzles swivelled to their extreme position to try to 'right' the rocket, causing it to break apart.

The investigation committee issued many recommendations.

— No software function should run during flight unless it is needed.

— Prepare a test facility including as much real equipment as technically feasible, inject realistic input data, and perform complete, closed-loop, system testing. Complete simulations must take place before any mission.

— Organize, for each item of equipment incorporating software, a specific software qualification review. Make all critical software a configuration controlled item.

— Review all flight software (including embedded software) and, in particular, identify all implicit assumptions made by the code and its justification documents on the values of quantities provided by the equipment. Check these assumptions against the restrictions on use of the equipment.

— Include participants external to the project when reviewing specifications, code, and justification documents. Make sure that these reviews consider the substance of arguments, rather than checking that verifications have been made.

— Give justification documents the same attention as code.

Many of these recommendations are applicable to accelerators. In this paper, we will discuss procedures, systems, and techniques to handle and mitigate the risks related to designing, deploying and operating machine protection systems.

## 1.1 Accelerator controls are complex systems

Control systems comprise many parts, and opportunities for malfunctioning are everywhere, e.g.:

— software fails unsafe;

— hardware fails unsafe;

- changes made on the wrong version of a program;
- wrong data received from sensors (but interpreted as true);
- a system was changed and cannot be brought back to a previous state;
- a system needs to be upgraded or changed, but there is not enough documentation to do it;
- system compromised by a malicious piece of code, which may go unnoticed for a long time;
- system hacked into.

Given this scenario, some risk mitigation strategies are:

- redundancy;
- life cycle management;
- fail-safe design;
- configuration control;
- quality assurance and quality control;
- standards;
- tests;
- documentation;
- cybersafety.

The concept of 'functional safety' is the corpus of concepts, processes, and guidelines that will enable us to mitigate those risks.

## 2 Functional safety

### 2.1 Introduction

For classic electrical and electronics based systems, there are three ways to improve safety: reduce component failure rate, increase diagnostics, and employ redundancy. Modern electronics, such as programmable logic controllers, microcontrollers, field programmable gate arrays and application-specific integrated circuits are powerful enough to be used to implement complex diagnostic schemes and control strategy reconfiguration on fault detection; in many cases, redundancy can be implemented with little cost increase. However, these features come at the cost of increased hardware complexity and introduction of software, which is more difficult to verify and validate for safety applications.

### 2.2 Life cycle management

A good life cycle management strategy is imperative for traceability and control. The standard V-model, shown in Fig. 1, represents a verification and validation model. Just like the waterfall model, the V-shaped life cycle is a sequential path of execution of processes: each phase must be completed before the next phase begins. Product testing is planned in parallel with a corresponding phase of development.

The model is advantageous in that it is simple to use, and such activities as planning and test design happen well before coding, saving time and increasing the chance of success (also, since defects are found at an early stage, they do not propagate quickly). Conversely, this approach might prove rigid, and requires software and hardware to be developed during the implementation phase. Moreover, if any changes happen mid-testing, then the test documents, along with requirement documents, must be updated.

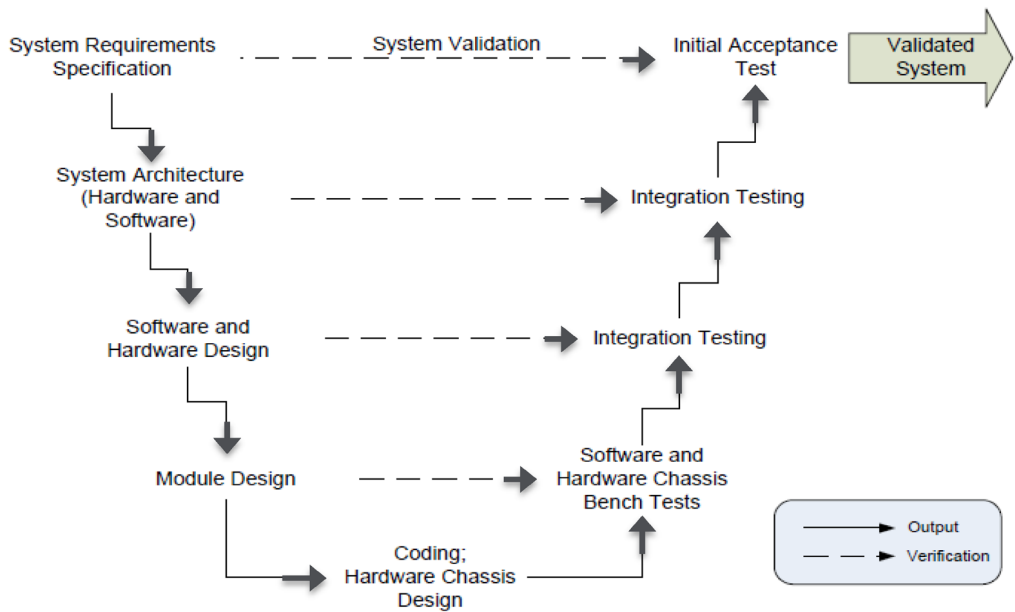

**Fig. 1:** V-model

### 2.3 Redundancy

Redundancy has different types and can be implemented at different levels. Sometimes it is two capacitors on a circuit-board, in case one fails; other times it is the duplication of a whole system, such as in some programmes of the 1950s, where redundancy was built into each and every component of an entire missile. The most common redundancy employed is parallel redundancy, where redundant parts, channels, or systems are active all the time. With a proper designed sensing and switching scheme, standby redundancy can also be employed.

The standards ISO 14118 *Safety of Machinery—Prevention of Unexpected Start-Up* [1] and IEC 60204-1 *Safety of Machinery, Electrical Equipment of Machines* [2] both state that reliance on a single-channel programmable electronic system is not recommended for safety. The IEC 60204-1 recommendation in particular is interpreted by many as an absolute ban on safety functions being implemented by programmable electronic systems in the sector.

IEC 61508 *Functional Safety of Electrical/Electronic/Programmable Electronic Safety-Related Systems* [3] has been published in recognition of the increasing use of this technology throughout a wide range of industrial uses.

Failures can be divided into two categories: common cause and common mode. Common cause failure is defined as one or more event causing concurrent failures of two or more separate channels in a multiple channel system, leading to system failure. Common mode failures are failures of two or more channels in the same way, causing the same erroneous result. Hardware redundancy is very effective in improving reliability: in systems employing redundancy, common cause or common mode failures usually dominate system-level failures. Compared with systems with identical duplicating components and circuits, systems with diversity redundancy (non-identical components) are less vulnerable to common cause and common mode failures.

To get the most out of redundancy, a managerial system is also required to determine, indicate, mediate, and isolate failures such that both safety and availability can be achieved (e.g. four engines on aircrafts).

## 2.4 Choosing components

Components from manufacturers with good quality-control systems and better manufacturing quality are preferred. It is expected that components with manufacturing defects, which contribute to early failure, have been identified and blocked by the quality-control system. Moreover, a better manufacturing quality will make the device less likely to fail during the normal working life, i.e., it will have a lower failure rate.

There are other approaches to improving component's reliability. Control of operating environment is the most important. The working environment, temperature, humidity, vibration, etc., should be controlled so that it matches the required working conditions for the electronics component. Thermal stress is another important factor that affects electronics reliability. Reducing electrical stress by lowering voltage or current also helps. When the hardware design is complete and ready for reliability predication, these factors will be required as inputs for evaluation.

## 2.5 Diagnostics and fail-safe design principle

If the detection of certain failures is feasible, the fail-safe design principle should be applied. This is particular useful if there is no way to tolerate fault consequence: 'fail-safe' design makes provisions for loss of energy source or control signal. Therefore, a 'de-energize-to-trip' philosophy is adopted in safety system design, so that system safety will not be jeopardized during power loss or absence of circuit integrity.

For complex systems where multiple failure modes exist, implementation of the fail-safe principle involves diagnostics: the system's integrity will depend on the information provided by the diagnostics to determine the nature of the failure and take corresponding action. With the aid of diagnostics, the failure of a component or a system can be classified as 'detected' or 'undetected'. For 'detected' failures with mild or safe implication, the user should be alerted, while for a 'detected dangerous' failure mode, the system should be brought to a safe state to ensure safety performance.

A fail-safe device or system is expected to fail eventually but, when it does, it will be in a safe way (for example, a ratchet mechanisms is used in lifts and elevators so that they cannot drop if the cable breaks). A fail-safe physical device may also define what occurs when a user error causes the system to behave in an undesired manner. In the case of software, there is no physical strain on systems, so the concept of mean time between failures is arguably inapplicable. However, software systems can and do fail all the time; for example, the following may happen:

— underlying hardware failure (e.g., networks and servers);

— external system failure (e.g., timing system failure);

— user error.

It is tempting to try to correct a failure situation and keep on running but this can lead to a system moving into an unknown state and creating more issues, as in these examples.

— The network is not responding but the system keeps on processing inputs and queuing outputs, expecting the network to respond later. Caches and disks fill up, affecting other systems, so, even when the network functionalities are restored, the system has to process hours' worth of data.

— A sensor seems to be showing the wrong data, but the system keeps running.

The solution is to institute limits on actions for recovery situations, e.g., by retrying only three times, setting a time limit on caches, etc. It is equally dangerous to make generic assumptions about correcting data across a system. If an input seems wrong, it is better to fail it, since one has no idea why the data do not seem to make sense, and the error is being hidden.

It is important not to simply put the system into a safe state, but also to inform those who can resolve the situation: error reporting and monitoring services should be designed upfront, and should define how operators should be kept informed.

## 2.6 Functional safety and safety integrity level

With the adoption of complex electronics into safety system applications, software must coordinate with hardware, such as microcontrollers. Although field programmable gate arrays can be purely hardware-only, with no run time software, the development process is very software-intensive, using complex software to design and verify the application. Therefore, verifying software is becoming a new challenge in safety system design, and the approach is that of 'functional safety'.

Another factor that contributes to the wide acceptance of functional safety comes with the adoption of a risk-based approach. Traditional descriptive standards and regulations list requirements for bottom-line protection; with a wide spectrum of applications, such an effort is increasingly difficult and requirements may be too conservative for some cases. A risk-based approach allows each application to carry out a risk assessment to determine the safety function and associated safety integrity level, such that there is no over-design or under-design.

The safety integrity level is the probability of a safety-related system performing the required safety function under all the stated conditions within a stated period of time (or put differently, the probability of failure on demand). 'Functional safety' standards originated from the IEC 61508 standard, and have spread to multiple applications, including process, machinery safeguarding, nuclear, and radiological industries.

The standard is sector independent in seven parts, the first four of which have been assigned basic safety publication status. This is the first international standard to quantify the safety performance of an electrical control system that can be expected by conforming to specified requirements, not only for the design concept but also for the management of the design process, operation, and maintenance of the system throughout its life cycle, from concept to decommissioning. These requirements, therefore, safely control failure to function resulting from both random hardware failure and systematic faults. Consequently, the standard represents a bold step, as a proactive approach to quantified, objective safety by design.

To categorize the safety integrity of a safety function, the probability of failure is considered—in effect, the inverse of the safety integrity level definition, looking at failure to perform rather than success. This is because it is easier to identify and quantify possible conditions and causes leading to failure of a safety function than it is to guarantee the desired action of a safety function when called upon.

The safety integrity level concept has emerged from the considerable effort invested in the safety of systems over the past two decades. Two factors have stood out as principal influences.

1) A move from the belief that a system can be either safe or unsafe, i.e., that safety is a binary attribute, to the acceptance that there is a continuum between absolute safety and certain catastrophe, and that this continuum is a scale of risk. This has led to an emphasis on risk analysis as an essential feature in the development of safety-related systems.

2) A huge increase in the use of software (and complex hardware, such as microprocessors) in the field of safety. This has led to a change in the balance between random and systematic faults. Previously, it was normal to assume (often implicitly) that safety could be achieved through reliability, and to deduce a value for the reliability of a system by aggregating, often through a fault tree, the random failure rates of its components. In some cases, failure rates were derived from historic use of the components and in others they were estimated, so the accuracy of the result was never beyond question. In fact, the greatest accuracy that could be achieved was that derivable from considering only random failures, for probabilistic methods are not valid for the analysis of systematic faults (those introduced, for example, through specification and design errors). With software, which does not wear out and in which all faults are systematic, there is no possibility of deducing system reliability by a method that is restricted to the consideration of random failures.

Another feature of software is its inherent complexity. Not only is it impossible to prove the absence of faults, but it would require an impracticably long time to derive high confidence in reliability from testing. So a number of problems arise for the developer, who needs not only to achieve but also to demonstrate safety.

The first consideration is that safety requirements may result from a risk analysis that may be quantitative or qualitative. However, as software failures result from systematic and not random faults, direct measurement of the probability of failure, or the probability of a dangerous failure, is not feasible, so qualitative risk analysis must be employed. While the reduction of a given risk may be defined as the specification of a software safety function, the tolerable failure rate of that function may be defined in terms of a safety integrity level. Depending on the standard in use, the safety integrity level may or may not be equated to numerical ranges of failure rates. Once risk analysis has led to a safety integrity level, this is used to define the rigour of the development process. The higher the safety integrity level, the greater the rigour, and tables are used in the standards to identify the methods, techniques, and management processes appropriate to the various safety integrity levels.

When a safety integrity level has been used to define the level of safety to be achieved, it follows that that safety integrity level should be the criterion against which a claim for the achieved safety is made (and judged). But if numerical values for the expected failure rate of software cannot be derived with confidence, it may not be possible to adduce proof of such a claim.

The IEC standard is based on a model relying on two entities: the equipment under control, which is used to provide some form of benefit or utility, and a complementary control system.

The standard recommends that the hazards posed by the equipment under control and its control system be identified and analysed and that a risk assessment be carried out. Each risk is then tested against tolerability criteria to determine whether it should be reduced. If risks are reduced by redesign of the equipment under control, we return to the starting point and hazard identification and analysis and risk assessment should again be carried out.

When it is decided that risk-reduction facilities should be provided in addition to the equipment under control and its control system, and that these should take the form of one or more electrical, electronic, or programmable electronic systems, then the terms of the standard apply to it or them.

The risks posed by the equipment under control and its control system may be contributed to by many hazards, and each must be mitigated until its risk is considered tolerable. The reduction of the risk associated with each hazard is specified as a 'safety requirement' and, according to the standard, each safety requirement must have two components: the functional requirement and the safety integrity requirement. The latter takes the form of a safety integrity level.

In Part 4 of IEC 61508, safety integrity is defined as "the likelihood of a safety-related system satisfactorily performing the required safety functions under all the stated conditions, within a stated period of time" and a safety integrity level as "a discrete level (one of 4) for specifying the safety integrity requirements of safety functions". Thus, a safety integrity level is a target probability of dangerous failure of a defined safety function.

The totality of the safety requirements for all hazards forms the safety requirements specification. Safety requirements are satisfied by the provision of safety functions, and in design these are implemented in 'safety-related systems'. The safety integrity levels of the safety requirements become those of the safety functions that will provide them, and then of the safety-related systems on which the safety functions are to be implemented. The separation of safety-related systems from the equipment under control and its control system (as by the provision of a protection system) is preferred. However, safety functions may also be incorporated into the control system and, when this is done, certain rules apply, to ensure that higher safety integrity level functions are not affected by the failures of lower safety integrity level functions.

Two classes of safety integrity level are identified, depending on the service provided by the safety function (Table 1):

— for safety functions that are activated when required (on demand mode), the probability of failure to perform correctly is given;

— for safety functions that are in place continuously (continuous mode), the probability of a dangerous failure is expressed in terms of a given period of time (per hour).

**Table 1:** Probability of failure

| Safety integrity level | Mode of operation: on demand (average probability of failure to perform design function on demand) | Mode of operation: continuous: (probability of dangerous failure per hour) |
|---|---|---|
| 4 | $\geq 10^{-5}$ to $<10^{-4}$ | $\geq 10^{-9}$ to $<10^{-8}$ |
| 3 | $\geq 10^{-4}$ to $<10^{-3}$ | $\geq 10^{-8}$ to $<10^{-7}$ |
| 2 | $\geq 10^{-3}$ to $<10^{-2}$ | $\geq 10^{-7}$ to $<10^{-6}$ |
| 1 | $\geq 10^{-2}$ to $<10^{-1}$ | $\geq 10^{-6}$ to $<10^{-5}$ |

The standard defines a low-demand mode of operation as 'no greater than one [demand] per year'. Since, in approximate terms, a year is taken to consist of $10^4$ hours, assuming a failure rate of once per year, the safety integrity level 4 requirement for the low-demand mode of operation is no more than one failure in 10 000 years. If there is to be no more than one demand per year made on a protection system, the equipment under control and its control system must have a dangerous failure rate of no more than once per year, or $10^{-4}$. However, arriving at this conclusion can be problematic because doing so is at the very limit of practical testability.

The failure rates attached to safety integrity levels for continuous operation are even more demanding (by a factor of $10^4$) and are intended to provide targets for developers. Because a system— certainly not a software-based system—cannot be shown to have met them, they are intended to define the rigour to be used in the development processes. Safety integrity level 1 demands basic sound engineering practices, such as adherence to a standard quality system, repeatable and systematically documented development processes, thorough verification and validation, documentation of all decisions, activities and results, and independent assessment. Higher safety integrity levels, in turn, demand this foundation plus further rigour.

The value of the safety integrity level lies in providing a target failure rate for the safety function or safety-related system. It places constraints on the processes used in system development, such that the higher the safety integrity level, the greater the rigour that must be applied. The processes defined as being appropriate to the various safety integrity levels are the result of value judgements regarding what needs to be done in support of a reasonable claim to have met a particular safety integrity level. However, the development processes used, however good, appropriate, and carefully adhered to, do not necessarily lead to the achievement of the defined safety integrity level. Even if, in a particular case, they did, the achievement could not be proved. But, even if evidence is insufficient to show that the safety integrity level requirement has been met, it does increase confidence in the system and its software.

Although the safety performance is the primary design objective, availability should also be considered. Large physics facilities are expensive investments; their productivity is critical financially and matters for the sake of science. Hence, there are always system-availability requirements for the project or the facility, and the availability of the indispensable safety system sets an upper bound for the whole facility's availability.

**2.7 Standards and guidance**

Standards are documents that establish uniform engineering and technical requirements for processes, procedures, practices, and methods. As this definition implies, some standards contain industry best practices, some provide description of interfaces such that interoperability can be achieved, while other standards simply describe methods for development and testing.

There are other, less formal, guidance documents that provide equally important information. They contain standard procedural, technical, engineering, or design information about the material, processes, practices, and methods covered or required by standards.

The definition of the term 'standard' includes the following:

— common and repeated use of rules, conditions, guidelines, or characteristics for products or related processes and production methods, and related management systems practices;

— definitions of terms; classifications of components; delineations of procedures; specification of dimensions, materials, performance, designs, or operations; measurements of quality and quantity in describing materials, processes, products, systems, services, or practices; test methods and sampling procedures; or descriptions of fit and measurements of size or strength.

We need standards to build our systems efficiently:

— deliverable products must be designed and built—they make use of procured items and must themselves be procured;

— each of these phases—procurement, especially—requires specification;

— effective specification requires standards.

An additional differentiation can be based on purpose:

— a basic standard has a wide-ranging effect in a particular field, such as a standard for metal, which affects a range of products from cars down to screws;

— terminology standards (or standardized nomenclature) define words, permitting representatives of an industry or parties to a transaction to use a common, clearly understood, language;

— test and measurement standards define the methods to be used to assess the performance or other characteristics of a product or process;

— product standards establish qualities or requirements for a product (or related group of products), to assure that it will serve its purpose effectively;

— process standards specify requirements to be met by a process, such as an assembly line operation, to function effectively;

— service standards, such as for repairing a car, establish requirements to be met in order to achieve the designated purpose effectively;

— interface standards, such as the point of connection between a telephone and a computer terminal, are concerned with the compatibility of products;

— standards on data to be provided contain lists of characteristics for which values or other data are to be stated for specifying the product, process or service.

International standards have been developed through a process that is open to participation by representatives of all interested countries, and that is transparent, consensus-based, and subject to due process.

Standards may also be classified by the intended user group, for example:

- organization standards are meant for use by a single industrial organization and are usually developed internally;
- industry standards are developed and promulgated by an industry for materials and products related to that industry;
- government standards are developed and promulgated by federal, state, and local agencies to address needs or applications peculiar to their missions and functions;
- international standards are developed and promulgated by international governmental and non-governmental organizations, such as the International Organization for Standardization (ISO);
- harmonized standards can be either an attempt by a country to make its standard compatible with an international, regional, or other standard, or it can be an agreement by two or more nations on the content and application of a standard, the latter of which tends to be mandatory.

### 2.7.1  *Software standards*

The ISO/IEC 12207 standard provides a common framework for developing and managing software. The IEEE/EIA 12207.0 standard consists of the clarifications, additions, and changes accepted by the Institute of Electrical and Electronics Engineers (IEEE) and the Electronic Industries Alliance (EIA), as formulated by a joint project of the two organizations. The IEEE/EIA 12207.0 standard outlines concepts and guidelines to foster better understanding and application of the standard. Thus, this standard provides industry with a basis for software practices that would be useable for both national and international business.

#### 2.7.1.1  *IEEE 12207—Software Life Cycle Processes*

This standard establishes a common framework for software life cycle processes, with well-defined terminology, that can be referenced by the software industry. It contains processes, activities, and tasks that are to be applied during the acquisition of a system that contains software, a stand-alone software product, or software service, as well as during the supply, development, operation, and maintenance of software products. Software includes the software portion of firmware. This standard also provides a process that can be employed for defining, controlling, and improving software life cycle processes.

The standard applies to the acquisition of systems and software products and services, to the supply, development, operation, and maintenance of software products, and to the software portion of firmware, whether performed internally or externally to an organization.

The standard groups the activities that may be performed during the life cycle of software into five primary processes, eight supporting processes, and four organizational processes. Each life cycle process is divided into a set of activities; each activity is further divided into a set of tasks.

In addition to IEEE12207, standard IEC61508, parts 3 and 7, focuses on safety functions with the following recommendations:

- use of structured and modular design;
- restricted use of asynchronous constructs;
- design for testability;
- restrictive use of ambiguous constructs;

— transparent and easy to use code;
— defensive code and range checking (to pick up faults or anomalies and respond in a pre-determined way);
— use of comments and annotations;
— limits on module sizes and number of ports to increase readability;
— avoidance of multi-dimensional arrays and go-to type commands;
— avoidance of redundant logic and feedback loops;
— avoidance of latches, asynchronous reset.

### *2.7.2   Hardware standards*

#### *2.7.2.1   Computer Automated Measurement and Control (CAMAC)*

This is a standard bus and modular crate electronics standard for data acquisition and controls, defined in 1972:

— solved the low-channel density problem of nuclear instrumentation methods;
— up to 24 modules in a crate, interfaced to a personal computer;
— not hot-swappable because of backplane design;
— data way management: module power, address bus, control bus, and data bus;
— 24-bit communication between controller and selected module.

#### *2.7.2.2   Versa Module Europa (VME)*

This standard backplane bus was defined in 1981:

— architecture not scalable for high speeds (single-ended parallel bus, not for gigabits per second);
— electromagnetic shielding not specified;
— developed for Motorola 68000 line of CPUs (the bus is equivalent to the pin of 68000 run out onto a backplane);
— faster bus (from 16 to 64 bit), up to 40 MHz (VME64).

#### *2.7.2.3   xTCA (Telecommunications Computing Architecture*

ATCA (Advanced Telecommunications Computing Architecture) and µTCA are platforms that provide:

— all-serial communications (multigigabits per second backplane);
— both complex experiment controls and large, high bandwidth and throughput data acquisition systems;
— the highest possible system performance, availability, and interoperability.

   To achieve high availability in a complex physics system, three main features are required:
1. modular architecture;
2. $N + 1$ or $N + M$ redundancy of single-point-of-failure modules (whose malfunction could stop operation of the machine or experiment);
3. intelligent platform management interface for quick isolation of faults and hot-swap.

In addition, physics modules need a few extended features:

— intelligent platform manager interface for cooling and thermal management, control and monitor;

— built-in hot-swap;

— designed for high-reliability;

— intelligent platform manager interface for cooling and thermal management, control, and monitor;

— built-in crate and component status monitoring and remote management and diagnostics;

— independent monitoring channel within the crate.

Specifically, µTCA is a modular, open standard for building high-performance switched-fabric computer systems in a small form-factor. At its core are standard advanced mezzanine cards, which provide processing and input–output functions. The µTCA standard was originally intended for smaller telecom systems at the edge of the network but has moved into many non-telecom applications, with standardized rugged versions becoming popular in mobile, military, telemetry, data acquisition, and avionics applications. The core specification, MTCA.0, defines the basic system, including backplane, card cage, cooling, power, and management. A variety of differently sized advanced mezzanine card modules are supported, allowing the system designer to use as much or as little computing and input–output as necessary. Subsidiary specifications (MTCA.1 to MTCA.4) define more rugged versions, specifically suited for military, aeronautic, and other demanding physical environments.

Modules (for a µTCA):

— cooling units;

— power modules;

— advanced mezzanine card for electronics, CPU, hard drives;

— rear transition module;

— µTCA central hub.

## 2.8 Tests

Testing is a process rather than a single activity, and starts as early as the system requirements specification. The choice of testing frequency definitely affects system reliability; the system design should accommodate such requirements, including setting up the test mode to facilitate testing. It is easy to see from the V-model that testing activities are a necessary step in completing every activity. Activities within the fundamental test process fall into the following basic steps (we will focus more on software tests, but the same principles apply to hardware tests):

1. planning and control;
2. analysis and design;
3. implementation and execution;
4. evaluating exit criteria and reporting;
5. test closure activities.

### 2.8.1 *Planning and control*

Test planning is intended to:

— determine the scope and risks and identify the objectives of testing;

— determine the test approach;

- implement the test policy or test strategy;
- determine the required test personnel and resources; test environments, hardware, etc.;
- schedule test analysis and design tasks, test implementation, execution, and evaluation;
- determine exit criteria.

A test strategy is created to inform project managers, testers, and developers of key issues of the testing process. This includes the testing objectives, method of testing, total time, and resources required for the project and the testing environments.

Test control is intended to:

- measure and analyse the results of reviews and testing;
- monitor and document progress, test coverage, and exit criteria;
- provide information on testing;
- initiate corrective actions;
- enable decision making.

### 2.8.2 *Analysis and design*

Test analysis and design should:

- review the test basis;
- identify test conditions;
- design the tests;
- evaluate testability of the requirements and system;
- design the test environment set-up and identify and required infrastructure and tools.

The test basis is the information needed to start the test analysis and create test cases. It is a documentation on which test cases are based, such as requirements, design specifications, product risk analysis, architecture, and interfaces. Test basis documents help understand what the system should do once built.

### 2.8.3 *Implementation and execution*

During test implementation and execution, test conditions are translated into test cases and procedures and scripts for automation, the test environment, and any other test infrastructure. (Test cases are a set of conditions under which a tester will determine whether an application is working correctly or not.)

Test implementation should:

- develop and prioritize test cases and create test data for those tests (to test a software application, for example, the tester needs to enter some data for testing most of the features: any such specifically identified data used in tests are known as test data);
- create test suites (a collection of test cases that are used to test a software program to show that it has some specified set of behaviours) from the test cases for efficient test execution;
- implement and verify the environment.

Test execution should:

- execute test suites and individual test cases, according to test procedures;
- re-execute tests that previously failed, to confirm a fix;

- log the outcome of test execution and record the identities and versions of the software under tests;
- compare actual results with expected results;
- report discrepancies between actual and expected results.

The test log is used for the audit trial. A test log records the test cases that were executed, in what order, who executed that test cases and the status of the test case (pass or fail).

### 2.8.4 Evaluating exit criteria and reporting

Based on the risk assessment of the project, criteria are set or each test level against which one can determine that 'enough testing' has been done. These criteria vary from project to project and are known as exit criteria.

Exit criteria are satisfied when:
- a maximum number of test cases are executed with a certain pass percentage;
- the software bug rate falls below a certain level.

### 2.8.5 Test closure activities

Test closure activities are performed when hardware or software is delivered, and include the following major tasks:
- check which planned deliverables are actually delivered and ensure that all incident reports have been resolved;
- finalize and archive test procedures, such as scripts or test environments, for later reuse;
- deliver test procedures to the maintenance organization;
- evaluate the testing process, to provide lessons for future releases and projects.

### 2.8.6 Proof tests

Safety system standards also require *proof tests*, to include verification of the following conditions:
- operation logic sequence given by cause and effect diagrams;
- operation of all input devices, including field sensors and single-instance storage input modules;
- logic associated with each input device;
- logic associated with combined inputs;
- trip set-point of all inputs;
- alarm functions;
- response time of the system (when applicable);
- functioning of manual actions bringing the process to its safe state, e.g., emergency stop;
- functioning of user-initiated diagnostics;
- safety system is operational after testing;
- all paths through redundant architectures should be tested.

Proof tests can identify 'hidden' device failures, although they cannot prevent failures from happening. The proof test interval should be large enough to catch failures, but too frequent testing also increases the likelihood of human errors in the system.

General considerations for proof testing are as follows:

— failure modes of the device and their effects on functionality; if a device failure is either self-revealed or can be detected by diagnostics, there is no need to include this device into proof testing;

— if a device has dominant age-related failure modes, preventive maintenance should be applied, in accordance with a reliability centred maintenance analysis;

— only safety-critical functions should be tested; non-safety-related functions should be included in another maintenance test or should simply be tested during an initial acceptance test and then again after a much longer interval;

— instruments and field devices that have no direct impact on safety, and are run under 'continuous mode' (continuous comparison among redundant devices), can either 'run to fail' or be tested with a much longer period (for example, signage);

— logic solvers that lack integrated diagnostic functions should still be tested annually, since they have no diagnostics and their functionalities can easily be changed;

— safety programmable logic controller-based logic solvers with strict management of change procedures have no need for a strict programmable logic controller-dedicated assurance test, however, the whole system should go through a full functional testing every 8–12 years (according to Shell standard DEP 32.80.10.10-Gen, July 2008);

— ease of testing should be considered during the system design stage, e.g., the process industry uses a 'maintenance override service' to facilitate online testing without tripping the process.

### 2.8.7  Test examples

As an example, these test methodologies can be followed for an accelerator safety system.

#### 2.8.7.1  Programmable logic controller or field programmable gate array bench test

This is part of the programmable logic controller or field programmable gate array software quality assurance activity; it demonstrates that the programmable logic controller field programmable gate array logic satisfies the specification.

#### 2.8.7.2  Interlock checks

These checks test field components subject to accidental damage or harsh environmental conditions, especially those of an electromechanical nature or which have moving parts; they are conducted at least every 6 months.

#### 2.8.7.3  Initial acceptance test

This is intended to test physical hardware, installations, all functions of a new safety installation (the installation includes hardware, programmable logic controller or field programmable gate array logic), including unintended functions and common mode failures, which could arise from design or implementation errors, or component malfunction. This test is carried out for a new installation or after major modifications. New safety software code downloads are currently considered a major modification.

#### 2.8.7.4  System features to be tested by initial acceptance test

These include:

— interaction between system and human–machine interfaces;

- each safety function, either loop-oriented or a complex functionality as a whole;
- degraded mode of operation if there are any requirements defined in operations requirements;
- recovery from failure;
- redundancy;
- different operation mode of the system;
- reasonable foreseeable abnormal conditions and misuse of the system.

*2.8.7.5 Safety assurance test*

This is intended to perform a maintenance function, to verify continuing operation of safety features; this test is carried out annually.

*2.8.7.6 Other tests*

Different industries and industrial standards use different terms for these test activities. For Safety Instrumented Systems, ANSI/ISA 84 (IEC 61511 Mod) contains requirements for factory acceptance testing, site acceptance tests, and proof testing. For complex process automation projects, IEC 62381 defines the scope and activities for factory acceptance testing, site acceptance tests, and site integration tests. Broadly speaking, the programmable logic controller bench test falls under the scope of factory acceptance testing; and the initial acceptance test is equivalent to a site acceptance test.

## 2.9 Configuration control

Configuration management is the unique identification, controlled storage, change control, and status reporting of selected intermediate components during the life of a system. Configuration control is the activity of managing the system and related document throughout the product's life cycle.

Configuration control ensures that:
- the latest approved version of the system and its components are used at all times;
- no change is made to the product baselines without authorization;
- there is a clear audit trail of all proposed, approved, or implemented changes.

When applied to software, there are additional challenges: on the one hand, individual developers need the flexibility to do creative work, to modify code to try out what-if scenarios, and to make mistakes, learn from them, and evolve better software solutions; on the other hand, teams need stability to allow code to be shared with confidence, to create builds and perform testing in a consistent environment, and to ship high-quality products with confidence. This requires an intricate balance to be maintained. Too much flexibility can result in problems, including unauthorized or unwanted changes, the inability to integrate software components, uncertainty about what needs to be tested and working programs that suddenly stop working. Conversely, enforcing too much stability can result in costly bureaucratic overhead and delays in delivery, and may even require developers to ignore the process in order to get their work done.

How is it possible to maintain the necessary balance between flexibility and stability, as software moves through the life cycle?

Some techniques include:
- selecting the appropriate type and level of control for each software artefact;
- selecting the right acquisition point for each configuration item;
- utilizing multiple-levels of formal control authority.

## 2.10 Quality assurance and quality control

Quality assurance is process oriented; quality control is product oriented: this might be one's starting point when considering how to assure quality to products and systems, as quality assurance makes sure that one is doing the right things, the right way, while quality control makes sure the results of what one has done are what one expected.

Assuring quality means more than making sure that quality exists also means stepping in wherever there are opportunities to add or ensure quality; for instance, clarifying requirements, documenting new requirements, facilitating communication among teams and, of course, testing. Testing should not be limited to hardware or software, but should extend to requirements, understanding of requirements, etc.

Typical quality assurance activities are: quality audit, defining process, selection of tools, and training.

Typical quality-control activities are: testing, walkthrough, inspection, and checkpoint review.

Any project should begin with a clear definition of requirements and deliverables. Performance requirements are defined in a 'black box' manner: the 'how' is not defined; the size, location, number of entry points, number and location of devices, entry requirements, desired access states, and operational modes must all be defined. Any interfaces to other systems should be highlighted and described in their own sections, for clarity.

The requirements document is carefully reviewed, since many future quality assurance tests reference these requirements; ideally, requirements documents should be maintained as 'living' documents.

A formal specification document is developed to define, specifically, how the system is to be built and operate. Specific parts are identified, system architectures of technology may be selected, input and output signal lists are defined. The systems specification should describe how to meet the requirements set forth in the requirements document. Any interfaces to other systems shall be highlighted and described in their own sections for clarity.

The specification document is carefully reviewed, since engineering and design work, as well as many future quality assurance tests, is performed against it.

### 2.10.1 *Technical reviews*

Technical reviews are conducted to evaluate design and engineering work for accuracy and performance against the requirements, specification, and other best-practice standards.

Informal peer reviews should be utilized periodically, to assess engineering work or testing procedures.

Formal reviews are conducted to evaluate project design and engineering work for accuracy and performance against the requirements, specification, and other best-practice standards. Formal reviews are typically specified in a project quality assurance plan. Large projects will typically have an early preliminary or system architecture review, followed by a detailed or final design review. At a minimum, there will always be at least one final technical review for a project. The membership of formal reviews follows a graded approach (Table 2). The number of external reviewers, the overall number of reviewers, and the organizational distance of reviewers from the overseeing organization is dependent on the scope, complexity, and technological familiarity of the proposed design compared to common practice.

**Table 2:** Minimum recommended reviewer complement

| | |
|---|---|
| Minor modification, familiar methods | 1 external reviewer |
| Minor modification, new methods | 2 external reviewers |
| Medium change, familiar methods | 2 external reviewers |
| Medium change, new methods | 2 or more external reviewers, 1 external to control department |
| Large change, familiar methods | 3 external reviewers, 1 external to control department |
| Large change, new methods | 3 or more external reviewers, 1 or more external to control department, 1 external to laboratory |

## 2.11 Documentation

Document management is the process of applying policies and rules to how documents are created, maintained, and archived within an organization. Document collaboration is merely the process of checking out, checking in, and versioning a document before it is published. Records management encompasses all of the functions of document management, but applies them to a broader set of content elements—not just documents.

The main aspects of managing a document through its life cycle include the following.

— Creation: Methods for envisioning, initiating, and collaborating on a new document's development.

— Location: There must be a physical location where documents will be stored and accessed. Usually, most documentation management systems require single-instance storage of a document so that there is only one version of the truth.

— Authentication and approval: Methods of ensuring that a document is fully vetted and approved before it is considered to be official compliant communication from the organization.

— Workflow: This describes the series of steps needed to pass documents from one person to another for various purposes, such as to gain approval to publish the document or to collect signatures on a document.

— Filing: For electronic systems, a document is filed by placing it in a physical location and then attaching metadata to the document. The metadata files the document logically by allowing the document to be found based on the metadata values assigned to the document.

— Distribution: Methods of getting the document into the hands of the intended readers.

— Retrieval: Methods used to find the documents, such as querying the index for keywords or using search alerts to find new content that meets the query keywords.

— Security: Methods used to ensure the document's integrity and security during its life cycle.

— Retention: Organization's policies and practices that inform everyone how long different document types are retained by the organization.

— Archiving: Similar in concept to retention, the differing characteristic is that archiving is a subset of retention policies. Archiving focuses on the long-term retention of documents in a readable format after the document's active life has ended. Subsumed in this category is the expiration of documents after they no longer need to be retained.

## 2.12 Cybersafety

Traditional network security risk management techniques are often inadequate to meet the specialized needs of control systems, whose security represents a unique challenge. Generally speaking, control systems are designed for accuracy, extreme environmental conditions, and real-time response in ways that are often incompatible with the latest cybersecurity technologies, inconsistent with consumer-grade hardware and software, and in conflict with common network protocols. As a result of these performance factors and limitations, engineers (rather than IT managers) have traditionally been responsible for the design, operation, and maintenance of control systems. Yet, despite their uniqueness, control systems are increasingly reliant on common network protocols, and connectivity often exists between control systems and enterprise networks, to include the Internet (Fig. 2).

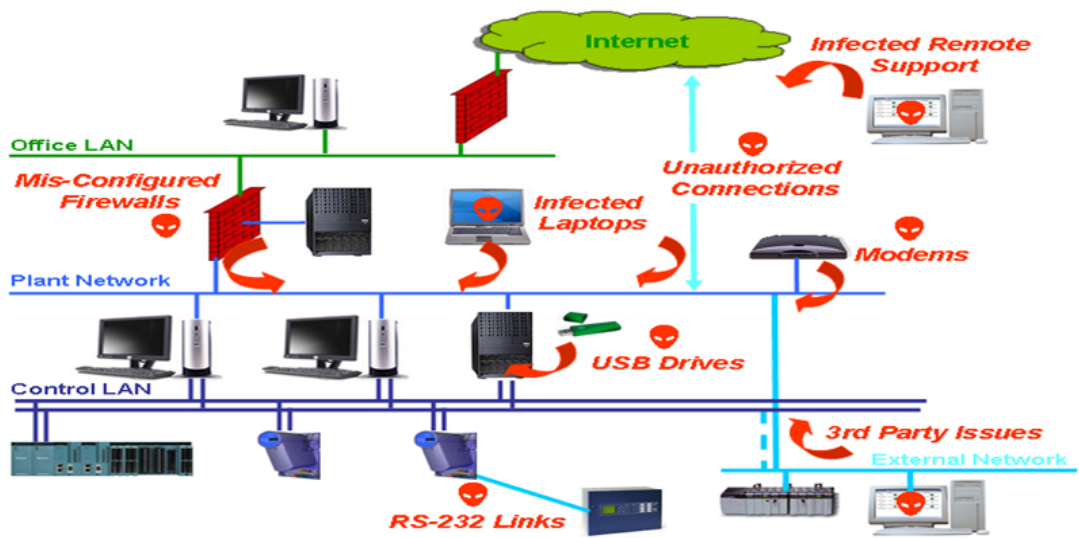

**Fig. 2:** Pathways into control systems

How does an organization ensure that its supervisory control and data acquisition system is secure? One of the answers is in standard ISA-99.02.01 (*Security for Industrial Automation and Control Systems: Establishing an Industrial Automation and Control Systems Security Program*), approved and published by the American National Standards Institute (ANSI). This readable standard lays out seven key steps for creating a cybersecurity management system for use with supervisory control and data acquisition and control systems.

The steps in ISA-99.02.01 are divided into three fundamental categories: risk analysis, addressing risk with the cybersecurity management system, and monitoring and improving the cybersecurity management system.

1. The first category lays out the stages an organization needs to follow to assess its current security situation and determine the security goals it wants to achieve.

2. The second category outlines processes to define security policy, security organization, and security awareness in the organization and provides recommendations for security countermeasures to improve supervisory control and data acquisition system security. The core idea in this section is a concept known as 'defence in depth', where security solutions are carefully layered to provide multiple hurdles to attackers and viruses.

3. The third category describes methods to make sure a supervisory control and data acquisition system not only stays in compliance with the cybersecurity management system but follows a continuous improvement programme.

## 2.12.1 Defence in depth

Sound strategy, regardless of whether it is for military security, physical security, or cybersecurity, relies on the concept of 'defence in depth'. Effective security is created by layering a number of security solutions so that if one is bypassed another will provide the defence. This means, for instance, not overrelying on any single technology, such as a firewall.

Defence in depth begins by creating a proper electronic perimeter around the supervisory control and data acquisition or control system and then hardening the devices within. The security perimeter for the control system is defined by both policy and technology. First, policy sets out what truly belongs in the control-system network and what is outside; next, a primary control-system firewall acts as the choke point for all traffic between the outside world and the control-system devices.

Once the electronic perimeter of the control system is secured, it is necessary to build the secondary layers of defence in the control system itself. Control components, such as human–machine interfaces and data historians based on traditional IT-operating systems, e.g., Windows and Linux, should take advantage of the proven IT strategies of patch and anti-virus management. However, this requires prior testing and care.

For such devices as programmable logic controllers and supervisory control and data acquisition controllers—where patching or anti-virus solutions are not readily available—industrial security appliances should be used. This solution deploys low-cost security modules directly in front of each group of control devices needing protection. The security modules then provide tailored security services, e.g., 'personal firewalling' and message encryption, to the otherwise unprotected control devices.

Table 3 compares the cybersafety requirements of a machine protection system with those of a personnel protection system (access control)—the latter being, generally, stricter.

## 2.13 Evolution of cybersafety landscape (a US perspective)

### 2.13.1 Framework for Improving Critical Infrastructure Cybersecurity (US National Institute of Standard, NIST, February 2014): a system of regulations and the means used to enforce them

The framework is based on:

— core functions (activities and references);
— implementation tiers (guidance);
— a framework profile (how to integrate cybersecurity functions within a cybersecurity plan).

The framework consists of four implementation tiers, each defined for three categories—risk management process, integrated risk management programme, and external participation. Any organization will follow into one of these three categories.

#### 2.13.1.1 Tier 1: Partial

— Risk management process: Organizational cybersecurity risk management practices are not formalized, and risk is managed in an ad-hoc and sometimes reactive manner.
— Integrated risk management programme: There is limited awareness of cybersecurity risk at the organizational level.
— External participation: An organization may not have processes in place to participate in coordination or collaboration with other entities.

**Table 3**: Cybersafety requirement comparisons

| Requirements | Personnel protection | Machine protection |
| --- | --- | --- |
| Use of configuration versioning system for software | Yes | Yes |
| Manage check-in and out of configuration versioning system with procedures | Yes | Yes |
| Track and check checksum | Yes; additional 'safety signature' available for safety-rated programmable logic controllers | Yes |
| Software download is password protected | Yes | Yes |
| Download over network? | No, not allowed; only local PROFIBUS (process field bus) connection allowed | Yes |
| Download to wrong CPU across network? | No; isolated networks and different CPU names and Internet protocol addresses even if on same network | No; isolated networks and different CPU names and Internet protocol addresses even if on same network |
| Protection against wrong safety program load | Hardware configuration is loaded; safety modules have hardware dual in-line package switches: hardware configuration error causes fail-safe shutdown | No |
| Physical isolation from controls network | No | No |
| Possible accidental (or act of sabotage) download of safety-critical code from controls network | No; local download only | Yes |
| Possible accidental changes (or act of sabotage) to supervisory control and data acquisition human–machine interface from controls network | Yes | Yes |

### 2.13.1.2 Tier 2: Risk informed

― Risk management process: Risk management practices are approved by management but may not be established as organizational-wide policy.

― Integrated risk management programme: There is an awareness of cybersecurity risk at the organizational level but an organization-wide approach to managing cybersecurity risk has not been established.

― External participation: The organization knows its role in the larger ecosystem, but has not formalized its capabilities to interact and share information externally.

### 2.13.1.3 Tier 3: Repeatable

― Risk management process: The organization's risk management practices are formally approved and expressed as policy.

― Integrated risk management programme: There is an organization-wide approach to managing cybersecurity risk.

— External participation: The organization understands its dependencies and partners and receives information from these partners that enables collaboration and risk-based management decisions within the organization in response to events.

#### 2.13.1.4 Tier 4: Adaptive

— Risk management process: The organization adapts its cybersecurity practices based on lessons learned and predictive indicators derived from previous and current cybersecurity activities.

— Integrated risk management programme: There is an organization-wide approach to managing cybersecurity risk that uses risk-informed policies, processes, and procedures to address potential cybersecurity events.

— External participation: The organization manages risk and actively shares information with partners to ensure that accurate, current information is being distributed.

### 2.13.2 NIST Special Publication (SP) 800-53 (Computer Security Guide)—Revision 4, April 2013

This standard is based on an information security programme: it covers risk assessment; policies and procedures; subordinate plans; training; periodic testing; incident response; and continuity of operations.

The standard is mission-oriented. It is based on FIPS 199 (Federal Information Processing Standard) for Security Categorization of Federal Information and Information Systems, and it includes definitions of security control categories for information systems (based on the key aims of confidentiality, integrity, availability).

The standard is also based on the impact on an organization's capability to accomplish its mission. (There is a full catalogue, including access control, awareness and training, audit and accountability, authentication, maintenance, media protection and access.)

### 2.13.3 Other standards

#### 2.13.3.1 IEC 17799: Information Technology—Code of Practice for Information Security Management

This standard is of a high level; being broad in scope and conceptual in nature, it forms a basis to develop customized security standard and security management practices.

#### 2.13.3.2 ISA-TR99: Integrating Electronic Security into the Manufacturing and Control System Environment

This standard is a guide to user and manufacturers. It can be used to analyse technologies and determine their applicability in securing manufacturing and controls.

#### 2.13.3.3 IEC 15408 (3.1): Information Security Management Systems (ISMS)

This standard provides a framework to specify security functional and assurance requirements through the use of protection profiles. Vendors can implement security attributes and testing laboratories can evaluate products.

#### 2.13.3.4 IEC 27001:2005: Common Criteria (CC) for Information Technology Security Evaluation

This is a system to bring information security under explicit management control through policies and governance; asset management; human resources security; access control; incident management; business continuity; etc.

*2.13.3.5 NIST SP 800-82*

This standard formalizes the defence-in-depth strategy: layering security mechanisms to minimize the impact to one mechanism as a result of failure.

The standard covers:

— Internet connection sharing policies based on Department of Homeland Security threat level;

— implementation of a multi-layer network topology;

— provision of logical separation between corporate and Internet connection sharing networks;

— use of a demilitarized zone (i.e., no direct communication between Internet connection sharing and corporate use);

— fault-tolerant design;

— redundancy for critical components;

— privilege management;

— encryption.

For laboratories, the risk tolerance for 'generic' Internet connection sharing is different than that for personnel protection or medical technology (the protection of lives, information, assets, etc.). Boundaries and interfaces have to be identified; moreover, in highly regulated environments, once a standard is chosen and committed, the organization can be audited against it.

# 3 An example: the Linac Coherent Light Source machine protection system

The machine protection system at the Linac Coherent Light Source (LCLS-I) at the SLAC National Accelerator Laboratory is an interlock system responsible for turning off or reducing the rate of the beam in response to fault conditions that might damage or cause unwanted activation of machine parts.

The system is required to:

— turn off or limit the rate of the electron beam when faults are detected, to prevent damage to sensitive machine components;

— protect undulator permanent magnets from the electron beam, limiting the radiation dosage to below a specified amount;

— protect beamline components from excessive beam exposure, to prevent damage to the vacuum system and unnecessary activation;

— shut off the beam (detect and mitigate) within one pulse at 120 Hz (i.e., 8.33 ms for LCLS-I);

— protect the laser heater system from the injector laser;

— allow fault conditions to set different maximum rates for each mitigation device;

— allow automatic beam rate recovery (after a fault is corrected, the beam rate is raised to its before-fault value);

— bypass faults securely;

— provide a user interface that quickly identifies system trips, allows 'post-mortem' analysis and shows history;

— provide the ability to change the configuration of the logic and beam rate by adding and removing input signals, bypassing device fault inputs, and setting and changing fault thresholds.

The machine protection system (see Fig. 3) is able to reduce the beam rate only to below the operators' requested beam rate, and cannot raise the beam rate above operators' requested beam rate. Separate systems support the machine protection system, to protect other energized devices such as power supplies, magnets, and klystrons. A separate beam containment system ensures that no beam or radiation reaches potentially occupied areas. To perform its functions, the machine protection system relies on a set of inputs and output signals (see Fig. 4).

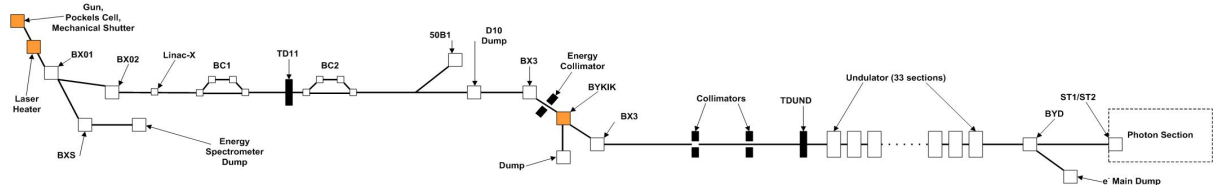

**Fig. 3:** Linac Coherent Light Source machine protection system

## 3.1 Inputs

### 3.1.1 Inputs from obstructions

Obstructions cover numerous devices, including vacuum valves, tune-up dumps, beam finder wires, and profile monitor screens. The beam is turned off whenever an obstruction reads a 'not-out' status. Beam finder wires, the tune-up dump, and profile monitor screens allow a maximum 10 Hz repetition rate once they are fully inserted. All obstruction devices require two limit switches to be incorporated in the design, to indicate fully in and fully out positions. An inconsistent status between the two switches is to be treated as a machine protection system fault. Obstructions are regarded as a pre-emptive fault, where the beam is turned off before it can cause any damage, e.g. to:

— profile monitor screens;
— collimator jaws;
— dechirper plates;
— beam stoppers.

### 3.1.2 Beam loss monitors

Beam loss is regarded as an actual fault with a requirement that the beam be shut off before the next pulse can be delivered; this means that the overall system must detect and mitigate the fault in less than 8.3 ms for 120 Hz operation.

Loss monitors with different types of sensitivity (e.g., toroids; protection ion chambers; optical fibre beam loss monitors) are deployed in different locations. The signal is gated to coincide with the beam arrival time and compared with a programmable threshold; it will indicate a fault if the threshold is exceeded. Two programmable threshold settings are required:

— exceeding the lower threshold can allow the beam rate to be lowered by the machine protection system;
— exceeding the higher threshold should cause the beam to be shut off and require the machine protection system to be reset manually.

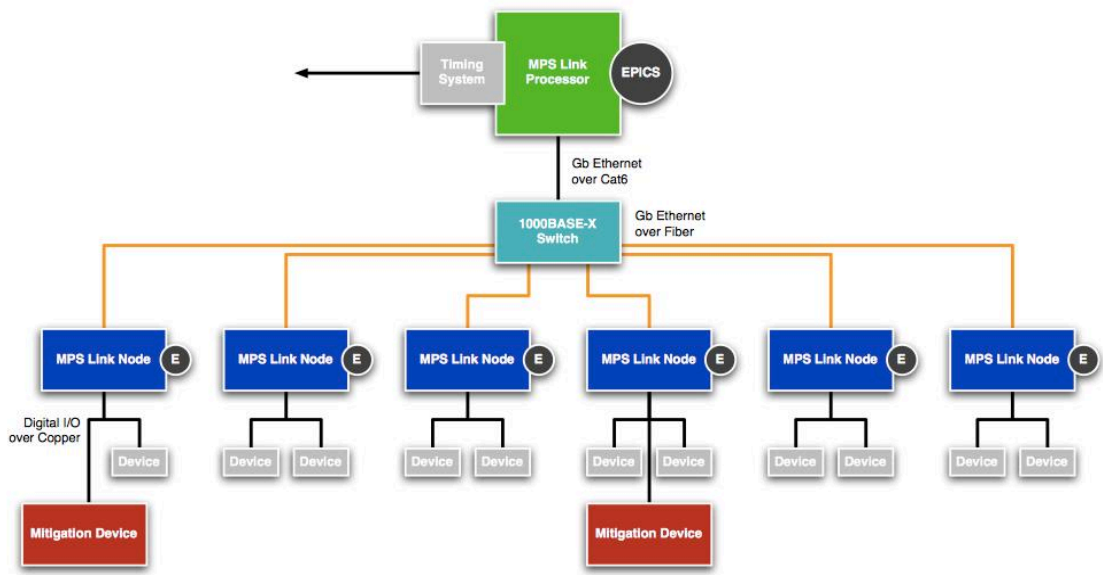

**Fig. 4**: Board diagram. EPICS, Experimental Physics & Industrial Control System; I/O, input–output; MPS, machine protection system.

### 3.1.3   Other inputs

These include:

- watchdog;
- vacuum valves;
- temperature readouts.

### 3.1.4   Sensors

There are a number of sensors. These include sensors for:

- vacuum valve position;
- water flow status;
- magnet power supply status;
- temperature;
- in-beam diagnostics status;
- beam position;
- beam charge;
- RF system status;
- beam containment status;
- beam loss.

## 3.2   Outputs (mitigation devices)

### 3.2.1   At the gun

The following devices interrupt the beam at the gun:

— laser heater mechanical shutter;

— photocathode laser mechanical shutter;

— gun trigger permit.

The mitigation scheme is based on shifting the timing of the gun RF from beam-time to standby time, while leaving the average rate of the laser and RF systems constant. The beam will not be extracted from the gun if the timing is shifted to standby time, since the RF is not present at the same time that the laser impinges on the cathode. The mechanical shutter is still deployed to block the light when the rate goes to zero. The shutter inhibits the injector UV laser before it hits the gun's cathode; its control is verified with optical position sensors; it faults the beam containment system mechanical shutter when the control does not match the position status.

### 3.2.2  *Pre-undulator fast kicker (BYKIK)*

The BYKIK (Fig. 5) is a pulsed dipole in LCLS-I, and is located in the middle of the DL-2 bend system in the linac-to-undulator beam line. The BYKIK is pulsed at a constant average frequency of 120 Hz and receives two input triggers, one at beam-time and the other at standby time. When the standby trigger is applied, the beam is transported unperturbed to the undulator. When the machine protection system shifts the trigger to beam-time, the beam is deflected by BYKIK onto a dump (collimator) and is not transported to the undulator. The switching of the BYKIK triggers between standby and beam-time can be done on a pulse-by-pulse basis so that either the beam is fully suppressed or bunches can be selectively allowed through to the undulator at a reduced rate. This feature is further exploited in special cases to send single shots and burst modes to the undulator on demand.

The secondary mitigation device requirements for BYKIK are derived from the need to suppress the beam to the undulator while the beam at the front end remains on at the full rate so that the machine is held stable by the beam-based feedback systems. For this to be reliable, the system must verify that BYKIK is operating correctly and dumping the beam before the undulator can be damaged. The verification is achieved at two levels. First, the BYKIK magnet control module signals a pre-emptive machine protection system fault if the magnet is out of tolerance within the specified time window of the pulse. The final verification must come from the beam itself at the time that BYKIK actually fires to kick the beam. For example, the beam position monitor immediately downstream of BYKIK should see the beam deflected by >1 mm, otherwise it should also register a machine protection system fault. In the event of either of these faults occurring, the beam is shut off at the gun.

### 3.3  Architecture

The system (Fig. 6) is based on a (dedicated, private) star network (Fig. 7) consisting of two entities: link processor and link nodes (interconnected over a private Gb ethernet network).

The machine protection system determines the maximum allowed beam rate by processing device fault input signals (from link nodes and input multiplexers) with a rate-limiting algorithm (executed on the link processor).

— The link node is the collection point of all sensor signals; it integrates sensor subsystems and drives mitigation devices.

— The link processor, in turn, runs the machine protection system control algorithm and makes decisions based on sensor states and interfaces to the timing system.

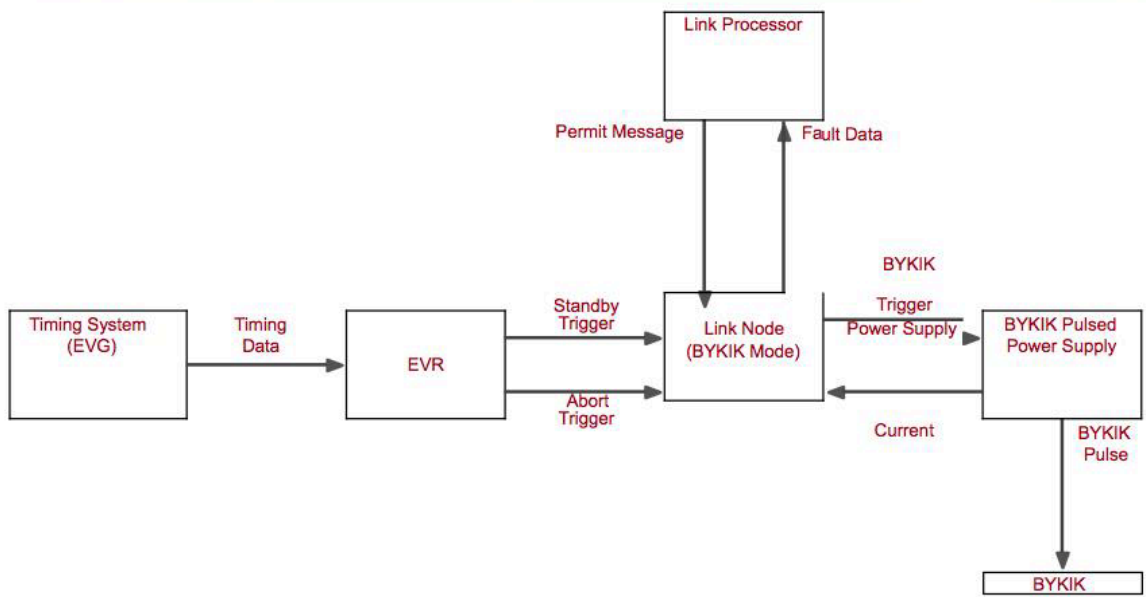

**Fig. 5**: BYKIK architecture: EVG, event generator; EVR, event receiver

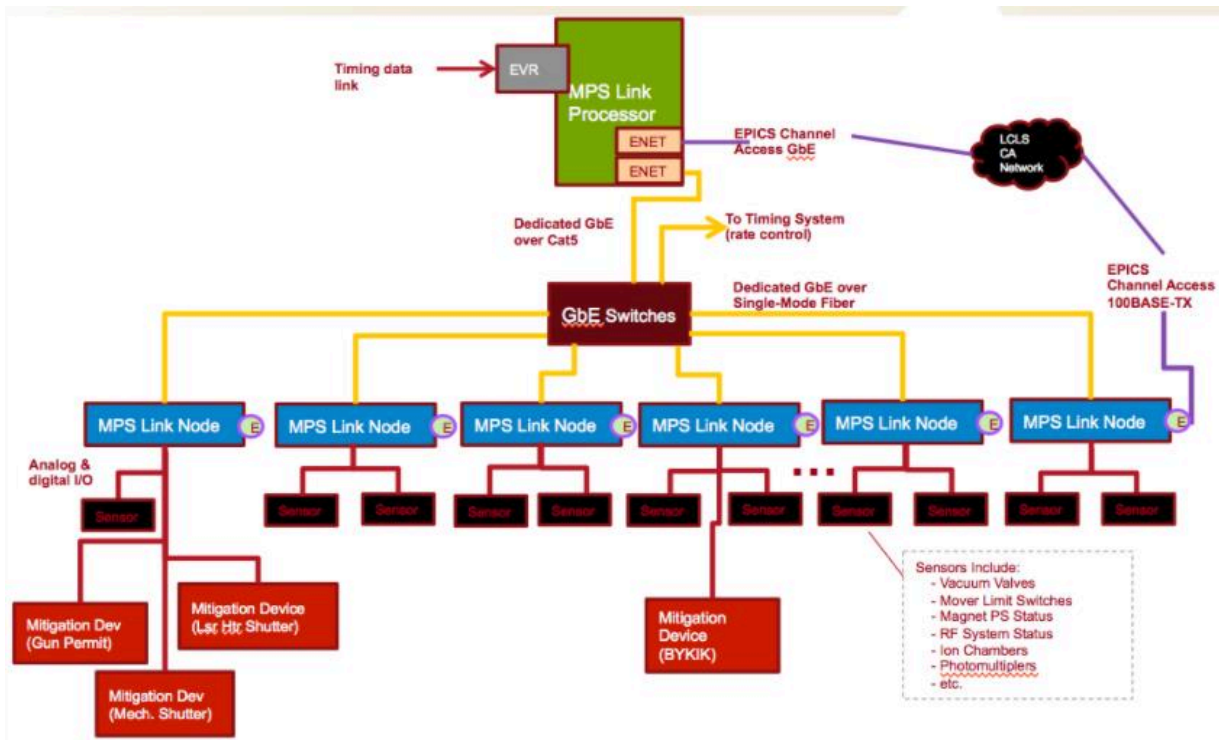

**Fig. 6**: Conceptual architecture diagram: CA, channel access; Dev, device; ENET, ethernet; EPICS, Experimental Physics & Industrial Control System; EVR, event receiver; GbE, Gb ethernet; LCLS, Linac Coherent Light Source; Lsr Htr, laser heater; Mech., mechanical; MPS, machine protection system; PS, power supply.

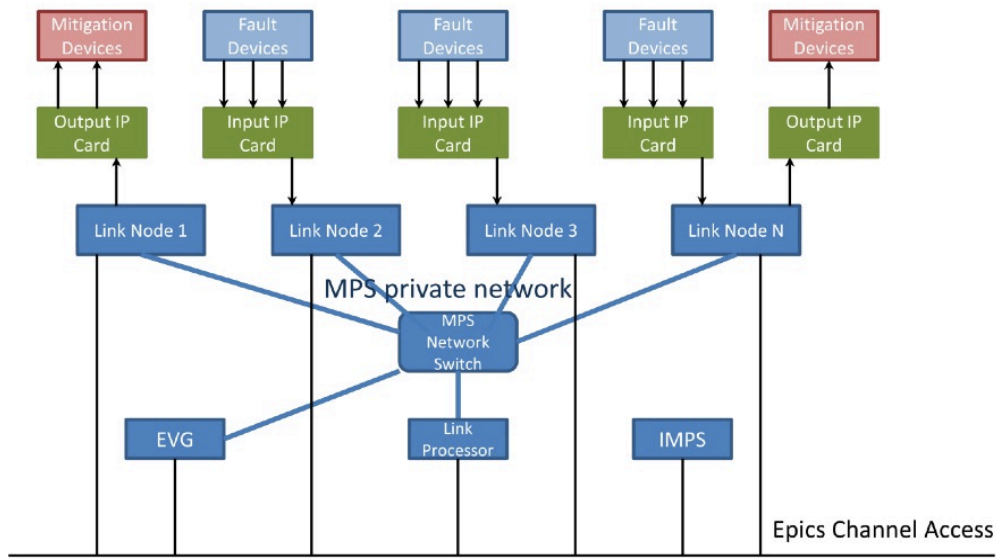

**Fig. 7**: Network architecture: EVG, event generator; IMPS, interface message processor system; IP, Internet protocol; MPS, machine protection system.

The link processor (a Motorola MVME 6100) has two copper Gb ethernet interfaces, a serial console port, and two peripheral component interconnect mezzanine card sites, along with an MPC7457 PowerPC processor that runs at 1.267 GHz, with 1 GB of RAM. The link processor's serial port is connected to a terminal server, a 1 Gb ethernet interface is used for high-speed communication with link nodes; the other is used for communication with the LCLS control system. It also sends synchronization and permit messages to link nodes. It faults all link node inputs to link nodes that provide a response within 8.3 ms.

The 32 LCLS link nodes are responsible for debouncing and latching digital inputs, digitizing analogue signals and comparing them with fault thresholds, and controlling the machine protection system mitigation devices. Link nodes are rack-mountable devices and occupy three rack units in a 19 inch rack. Built around the Xilinx Vertex four-field programmable gate arrays, each link node can be configured to support up to 96 digital inputs, 8 solid-state relay outputs, 4 TTL-compatible logic level trigger inputs, and 4 trigger outputs. One of each link node's two small form-factor pluggable slots is filled with a fibre-optic transceiver for high-speed communication with the link processor over the Gb ethernet. A full speed USB 1.1 port provides serial communication with the field programmable gate array while a separate DE-9 serial port gives access to the link node's EPICS (Experimental Physics & Industrial Control System) input–output controller serial port. The input–output controller serial ports are connected to terminal servers.

Four interface board slots allow signal conditioning to be placed between incoming signals and the link nodes' Industry Pack cards. Commercial off-the-shelf analogue-to-digital converter and digital-to-analogue converter Industry Pack cards are used to control and read back beam loss monitor high-voltage power supply voltages. A charge-integrating analogue-to-digital converter (QADC) Industry Pack card is used to digitize up to eight protection ion chamber or beam loss monitor signals, allowing each link node to monitor up to 32 analogue signals. The digitized signals are compared in the link node field programmable gate array against thresholds set by the link node's input–output controller via EPICS. Only the Boolean results of these comparisons are sent to the link processor for fault mitigation.

### 3.4 Communication

All time-critical data are transmitted over the machine protection system's dedicated Gb ethernet network using the user datagram or Internet protocol. The link processor uses a real-time protocol stack originally created for the LCLS beam position monitor data acquisition system. The real-time protocol stack not only provides deterministic behaviour for the messaging, but also allows ordinary network hardware and software tools to be used to build and test the system, since no new protocols are introduced. On the link node side, the network stack is implemented in the field programmable gate array firmware. A stack of dedicated Gb ethernet switches connects the link nodes and the link processor These switches queue and serialize concurrent data sent to the link processor and also handle the physical layer conversion of the link processor's copper and the link nodes' fibre Gb ethernet connections.

When the link processor is woken by the 360 Hz signal from the LCLS timing system, it broadcasts a synchronization message to all link nodes, requesting updated fault data, and providing the timing system's newest time-stamp. In response, the link nodes send the link processor a time-stamped status message containing all unacknowledged machine protection system device faults that have occurred since the previous synchronization message. The link processor copies the fault data to local buffers and returns the status message to the source link node. The link node uses this message as an acknowledgement of the faults that the link processor has received. All faults are latched in the link nodes and are cleared only when the link processor has acknowledged the fault and the fault itself has been cleared.

The link processor processes the faults using the currently running machine protection system logic and broadcasts a permit message to the link nodes. Link nodes allow the beam past their connected mitigation devices for 1/360 s if permitted. If a permit message is not received or if the beam is not permitted, link nodes stop the beam at their mitigation devices.

### 3.4.1 *History server*

The link processor logs all fault and status messages to a machine protection system history server application, which stores the messages in an Oracle database in real time. Machine protection system messages are stored separately from the normal logging system so that no messages are lost; they are also forwarded to the normal message logging system, so that they can be correlated with other logged events. A machine protection system history viewer is available to the operators via the machine protection system graphic user interface.

The events logged are:

— device state changed;

— beam rate changed;

— destination changed;

— history servers notify link processor of their existence.

### 3.4.2 *Faults bypass*

Device faults can be bypassed via an EPICS display by selecting a fault, choosing its bypass state, and supplying the bypass duration.

For example, an operator can choose to bypass a flow switch for one day by selecting the flow switch input, selecting its OK state, and giving a bypass duration of 24 h. All bypasses are logged and automatically timed by the machine protection system. The operator is alerted when the bypass time is reached, forcing the operator to re-evaluate bypasses.

### 3.5 Human–machine interface

The human–machine interface is illustrated in Figs. 8 to 11.

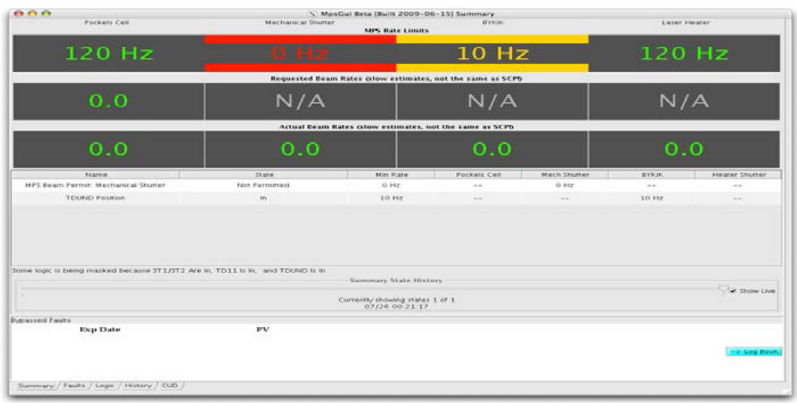
**Fig. 8**: MAIL machine protection system: graphic user interface

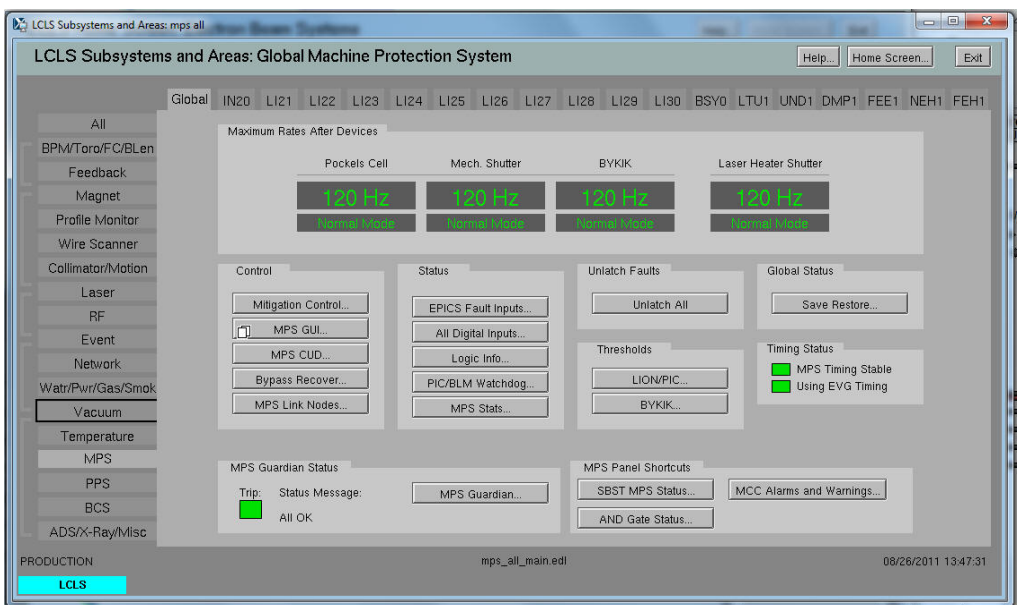
**Fig. 9:** Machine protection system: global panel

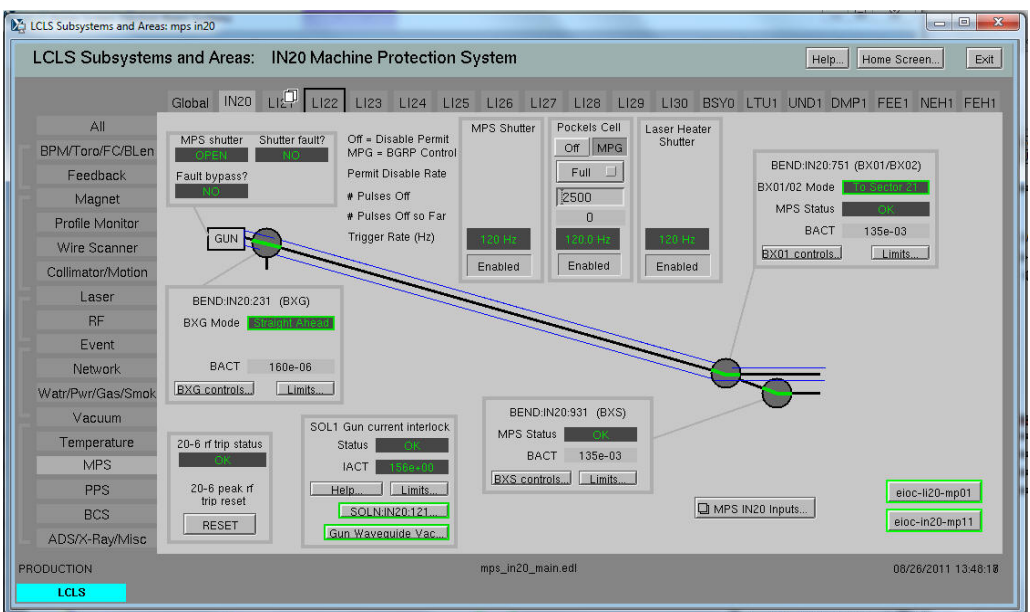
**Fig. 10:** Injector panel

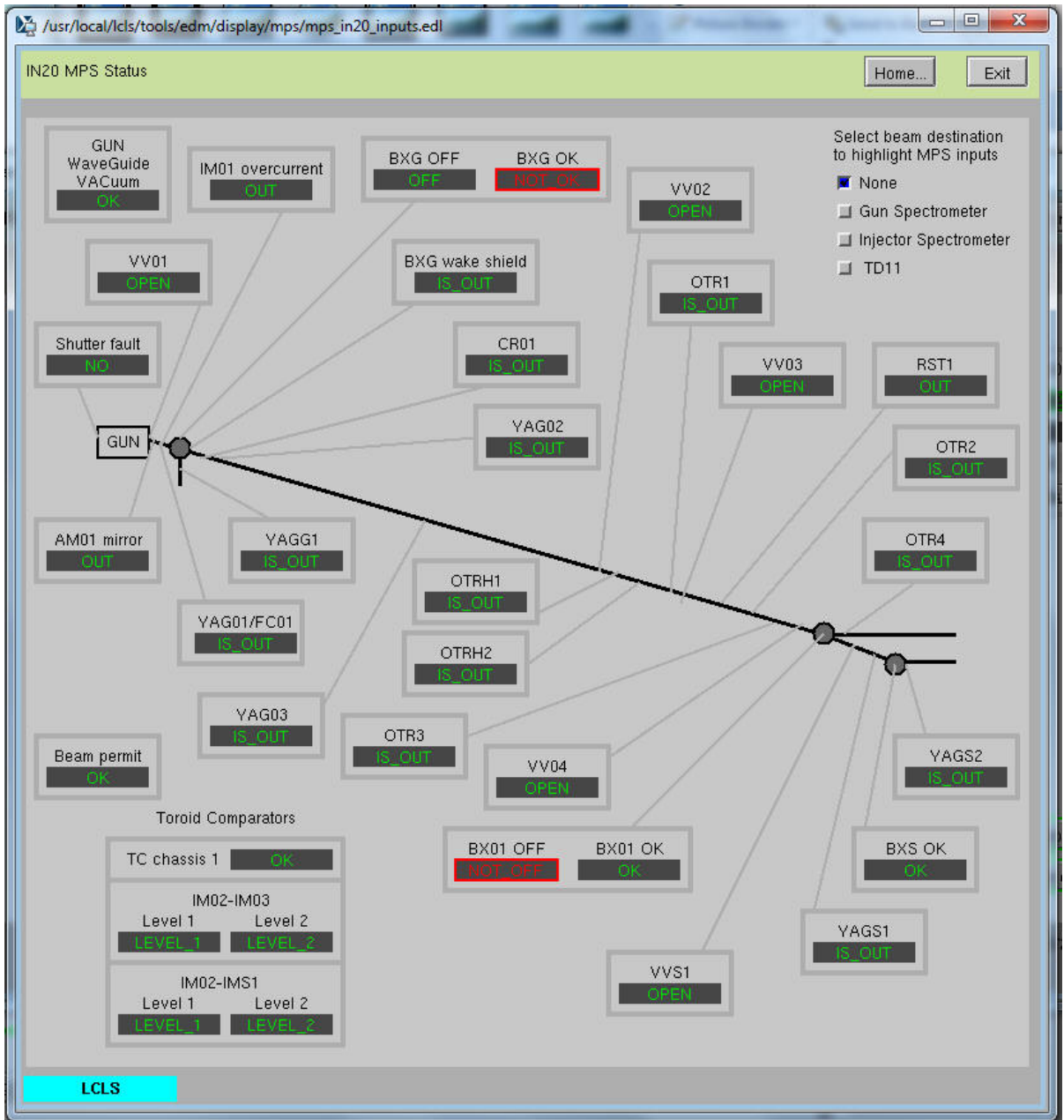

**Fig. 11:** Injector inputs

## Acknowledgements

The author wishes to thank M. Boyes and F. Tao of SLAC for many fruitful conversations and their insights on machine protection systems for accelerators.